# Large-Scale Conformal Growth of Atomic-Thick MoS$_2$ for Highly Efficient Photocurrent Generation

Tri Khoa Nguyen,[1] Anh Duc Nguyen,[1] Chinh Tam Le,[1] Farman Ullah,[1] Kyo-in Koo,[2] Eunah Kim,[3] Dong-Wook Kim,[3]* Joon I. Jang,[4]* Yong Soo Kim[1]*

[1]*Department of Physics and Energy Harvest Storage Research Center (EHSRC), University of Ulsan, Ulsan 44610, South Korea*

[2]*Department of Biomedical Engineering, University of Ulsan, Ulsan 44610, South Korea*

[3]*Department of Physics, Ewha Womans University, Seoul 03760, South Korea*

[4]*Department of Physics, Sogang University, Seoul 04107, South Korea*

*Corresponding authors: yskim2@ulsan.ac.kr (Y.S. Kim), dwkim@ewha.ac.kr (D.–W. Kim), and jjcoupling@sogang.ac.kr (J.I. Jang)

## Abstract

Controlling the interconnection of neighboring seeds (nanoflakes) to full coverage of the textured substrate is the main challenge for the large-scale conformal growth of atomic-thick transition metal dichalcogenides by chemical vapor deposition. Herein, we report on a controllable method for the conformal growth of monolayer MoS$_2$ on not only planar but also micro- and nano-rugged SiO$_2$/Si substrates via metal-organic chemical vapor deposition. The continuity of monolayer MoS$_2$ on the rugged surface is evidenced by scanning electron microscopy, cross-section high-resolution transmission electron microscopy, photoluminescence (PL) mapping, and Raman mapping. Interestingly, the photo-responsivity (~254.5 mA/W) of as-grown MoS$_2$ on the nano-rugged substrate exhibits 59 times higher than that of the planar sample (4.3 mA/W) under a small applied bias of 0.1 V. This value is record high when compared with all previous MoS$_2$- based photocurrent generation under low or zero bias. Such a large enhancement in the photo-responsivity arises from a large active area for light-matter interaction and local strain for PL quenching, where the latter effect is the key factor and unique in the conformally grown monolayer on the nano-rugged surface. The result is a step toward the batch fabrication of modern atomic-thick optoelectronic devices.

Large-scale synthesis of a two-dimensional (2D) semiconductor on an insulating substrate without any transfer process has recently drawn much attention to benefit the batch fabrication of ultra-thin optoelectronic devices such as photodetectors and solar cells.[1-5] Especially, monolayer $MoS_2$ is an ideal platform for realizing such devices owing to the direct bandgap of 1.8 eV and ultrafast carrier dynamics therein.[6-9] For instance, Yin *et al*. firstly reported on photocurrent generation from monolayer $MoS_2$-based phototransistor by applying a constant drain or gate voltage.[10] This work triggered numerous follow-up studies to improve photocurrent generation in monolayer $MoS_2$ by using external bias voltage, gas adsorbates, heterostructural contact, and photo-thermoelectric effects.[11-19] On the other hand, increasing photon absorbance by covering atomic-thick $MoS_2$ on a surface with engineered morphology would strongly improve photocurrent generation. Recently, Cho *et al*. reported on the significant enhancement of broadband photocurrent generation in $MoS_2$ decorated on the nanocone surface by pre-coated $MoO_3$ thin-film sulfurization.[20] In this case, however, the effective working area is only confined to triple layers $MoS_2$ decorated on the flat region between the nanocones, thereby limiting the magnitude of photocurrent. Moreover, light absorption is not optimized because of the indirect gap nature of tri-layer $MoS_2$. Clearly, it is important to establish a synthesis method to produce monolayer $MoS_2$ on a large effective area, although reaction sources are rather limited in thermal chemical vapor deposition (CVD).

An alternative solution is to conformally grow monolayer $MoS_2$ onto rugged substrates. Rugged substrates not only offer more active areas but also increase light mater interaction through their improved light trapping capabilities, which can eventually enhance photocurrent generation in an optoelectronic device. However, it is highly unlikely to acquire a conformally grown monolayer $MoS_2$ film over a rugged substrate via a conventional CVD technique using solid-phase precursors such as $MoO_3$, $MoCl_5$, and S.[20-22] The key challenge is to maintain the density of precursors in a CVD reaction chamber uniformly during the reaction. In this regard, metal-organic precursors such as molybdenum hexacarbonyl ($Mo(CO)_6$) and diethyl sulfide (($C_2H_5)_2S$) can be suitable alternatives to produce uniform monolayer $MoS_2$ on various rugged substrates due to their high equilibrium vapor pressure near room temperature.

In this article, we report a record enhancement in photocurrent generation in the conformal monolayer $MoS_2$ on planar, micro-rugged, and nano-rugged surfaces grown by metal-organic chemical vapor deposition (MOCVD). Surprisingly, monolayer $MoS_2$ on the nano-rugged surface shows 59 times higher photocurrent compared to the case on the planar surface. The multifold increase in photocurrent generation could be understood in terms of (*i*) enhanced optical absorbance due to large active area and improve light-matter interaction and (*ii*) local strain that induces significant photoluminescence (PL) quenching. The latter effect is especially unique in our approach that arises basically from suppression of exciton recombination upon the band structure modification, thereby boosting the collection of charge carriers. Our study implies that MOCVD is a feasible route to ensure the large-scale conformal growth of monolayer $MoS_2$, which is important for realizing atomically thin, high-performance optoelectronic devices.

In order to synthesize conformal monolayer $MoS_2$ on a large scale substrate, MOCVD was employed using molybdenum hexacarbonyl ($Mo(CO)_6$) and diethyl sulfide (($C_2H_5)_2S$) as the precursors.[1] We found that a slow reaction rate was a crucial factor for the conformal growth of monolayer $MoS_2$, in addition to the flow rate of precursors, hydrogen (reduction gas) and Ar (carrier gas). More details can be found in **Experimental Section** and **Figure S1**. Hence, the long reaction time of 8 hours was set to complete the formation of monolayer $MoS_2$. The growth mechanism can be monitored by reaction-time-dependent morphologies of as-grown $MoS_2$ from ~5 hours to ~8 hours as shown in **Figure S2**. For a sufficient reaction time (8 hours or so), continuous monolayer $MoS_2$ is formed by the interconnection of neighboring nanoscale monolayer flakes. This argument is well supported by morphological images as depicted in **Figure S3**. Some of the neighboring flakes were found to be linked and no second-layer nucleation was observed before forming continuous $MoS_2$.

Furthermore, we found that our growth method can be directly employed for the conformal growth of monolayers $MoS_2$, even when the 300-nm-$SiO_2$/Si substrates has nonplanar fine structures. For example, **Figures 1(a)–(c)** show monolayer $MoS_2$, uniformly prepared on the micro-rugged surface. The corresponding cross-section high-resolution transmission electron microscopy (HR-TEM) images in **Figures 1(d)-(f)** were taken at the corner region of the micro-rugged surface, confirming the conformal growth of monolayer $MoS_2$ interconnected over the structures. Similar to the growth mechanism in the planar surface, prior to the formation of continuous monolayer $MoS_2$, nano-flakes located on the micro-rugged surface tend to merge together when the reaction time increases (**Figure S4**). The nature of grain boundaries formed at the sharp corner of the rugged structure is clearly revealed by the cross-section HR-TEM images [**Figures 1(g)-(i)**]. The thickness and width of the overlapped region of two nearby grains are ~2 nm and ~40 nm, respectively. The growth of MoS$_2$ on the nano-rugged surface also shows a similar manner as presented in **Figure S5**.

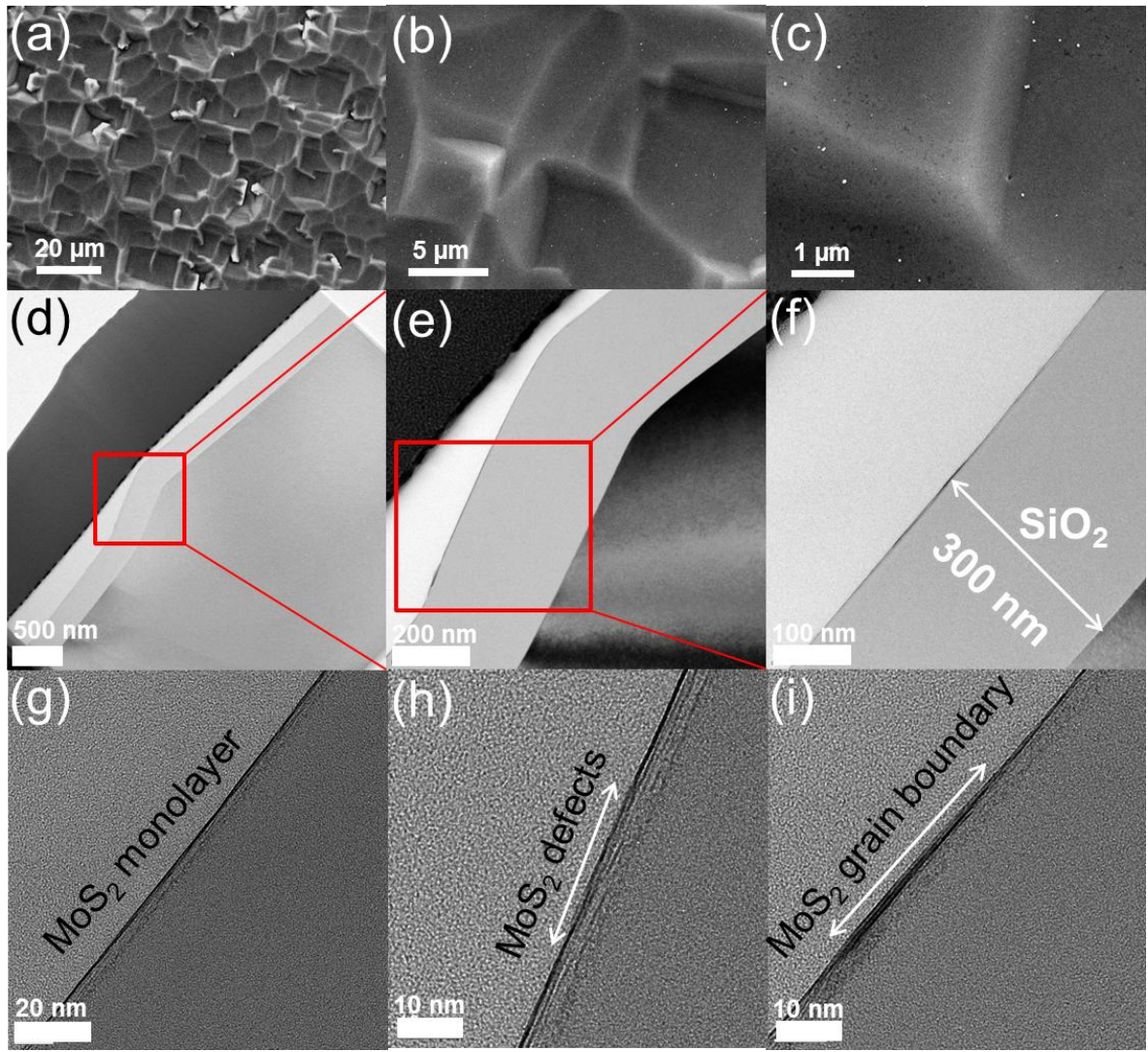

**Figure 1 (a)-(c) FE-SEM and (d)-(f) HR-TEM images of continuous monolayer MoS$_2$ conformally grown on the micro-rugged surface. (g)-(f) Cross-section HR-TEM images, showing the quality of the conformal monolayer on a nanoscopic scale.**

**Figures 2(a)-(c)** are FE-SEM images, depicting the morphologies of as-grown $MoS_2$ on the planar, micro-rugged, and nano-rugged substrates. The $MoS_2$ film on the planar substrate can be clearly seen based on the slight difference in the optical contrast between grain boundaries and grains. However, it is rather ambiguous to tell the presence of the $MoS_2$ films formed on the other substrates using microscopic imaging due to smaller grain sizes fully covering the rugged surfaces. Therefore, we carried out PL mapping using an excitation wavelength of ~473 nm with a sufficiently fine scan step (~0.4 μm), which is the necessary step to confirm the formation of the $MoS_2$ films on distinct terraces. The area of 20 $μm^2$ was chosen to conduct PL mapping. **Figures 2(d)-(f)** correspond to the PL mapping data arising predominantly from radiative recombination of *A*-excitons across the direct bandgap in our samples of different morphologies. Interestingly, we found that the PL peak position varies depending on the underneath terraces. In particular, this PL peak roughly centers at ~668 nm for the flat surface of $SiO_2$/Si substrate, but it is utterly blue-shifted to the wavelength of ~663 nm for the patterned surface in the nano-rugged specimen. For the micro-rugged sample, the mapping data reflect the gradual distribution of PL wavelengths in between the two extremes. While PL peaks are prominently located at the wavelength of 663 nm on the inward bending surface, they are scattered toward longer wavelengths on more localized planar portions without curvature. This wide distribution in the PL peak position can be interpreted by local strain induced by underneath scaffolds.[23] Analogous to artificially strained $MoS_2$, the $MoS_2$ films grown on the curved surfaces may undergo local strain with various amplitude and sign depending upon the underlying structures. This strain significantly impacts on the electronic band structure of $MoS_2$, which could decrease/increase the direct bandgap energy upon homogeneous tensile/compressive strain.[23-25]

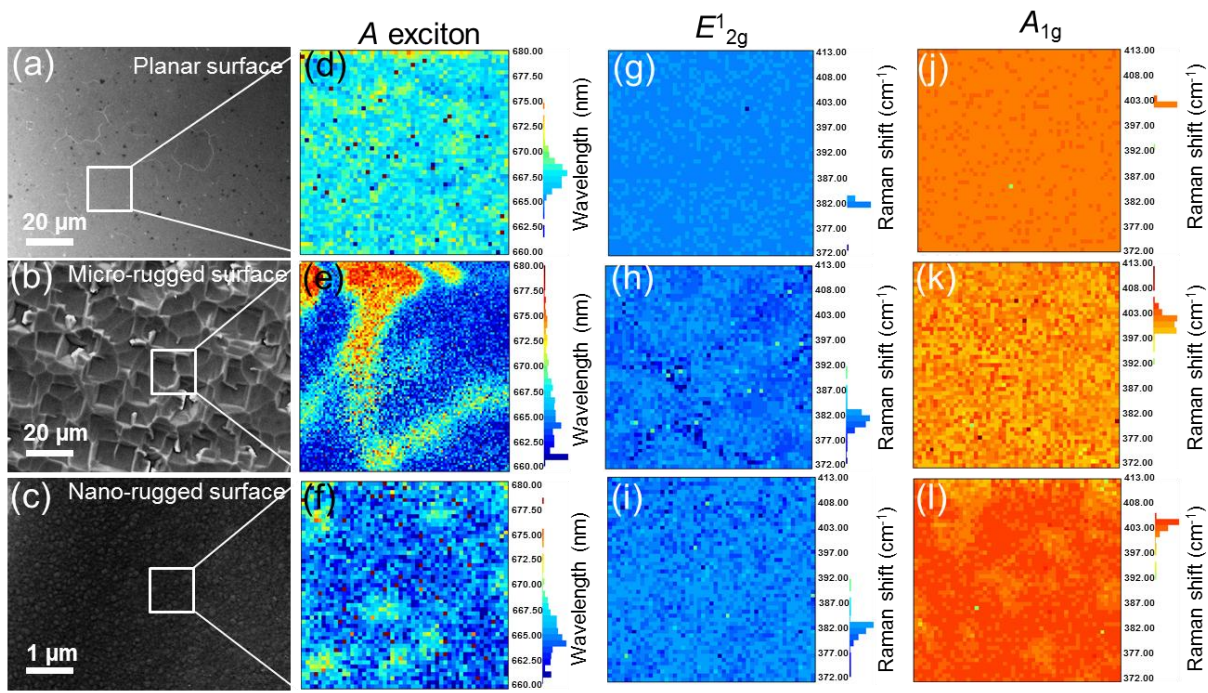

**Figure 2.** (a)-(c) FE-SEM images, (d)-(f) PL mapping of A-exciton emission, (g)-(i) Raman mapping of the $E^1_{2g}$ vibration mode, and (j)-(l) Raman mapping of the $A_{1g}$ vibration mode of MoS$_2$ covered on the planar, micro-rugged, and nano-rugged surfaces. The PL and Raman mapping were taken over area of 20 µm$^2$ using excitation wavelength of 473 nm with a scan step of 0.4 µm$^2$.

In order to examine the degree of conformal coating of our monolayer MoS$_2$, we mapped two characteristic phonon modes of MoS$_2$; in-plane $E^1_{2g}$ as shown in **Figure 2(g)-(i)** and out-of-plane $A_{1g}$ as shown in **Figure 2(j)-(l)**. The peak separations are determined to be 20 - 21 cm$^{-1}$, close to the previously reported values of monolayer MoS$_2$.[23-25] Together with HR-TEM analysis, the Raman mapping data indicate highly conformal coating of our films. These two phonon modes are almost unchanged in their vibrational frequencies over the scan area for the planar sample. In contrast, both of the $E^1_{2g}$ and $A_{1g}$ modes of the rugged surfaces are shifted ~2

cm$^{-1}$.[25] This slight shift in frequency infers that strain formed in MoS$_2$ on the rugged surface affects phonon vibrations, which is also consistent with the PL mapping results (**Figure S6**).

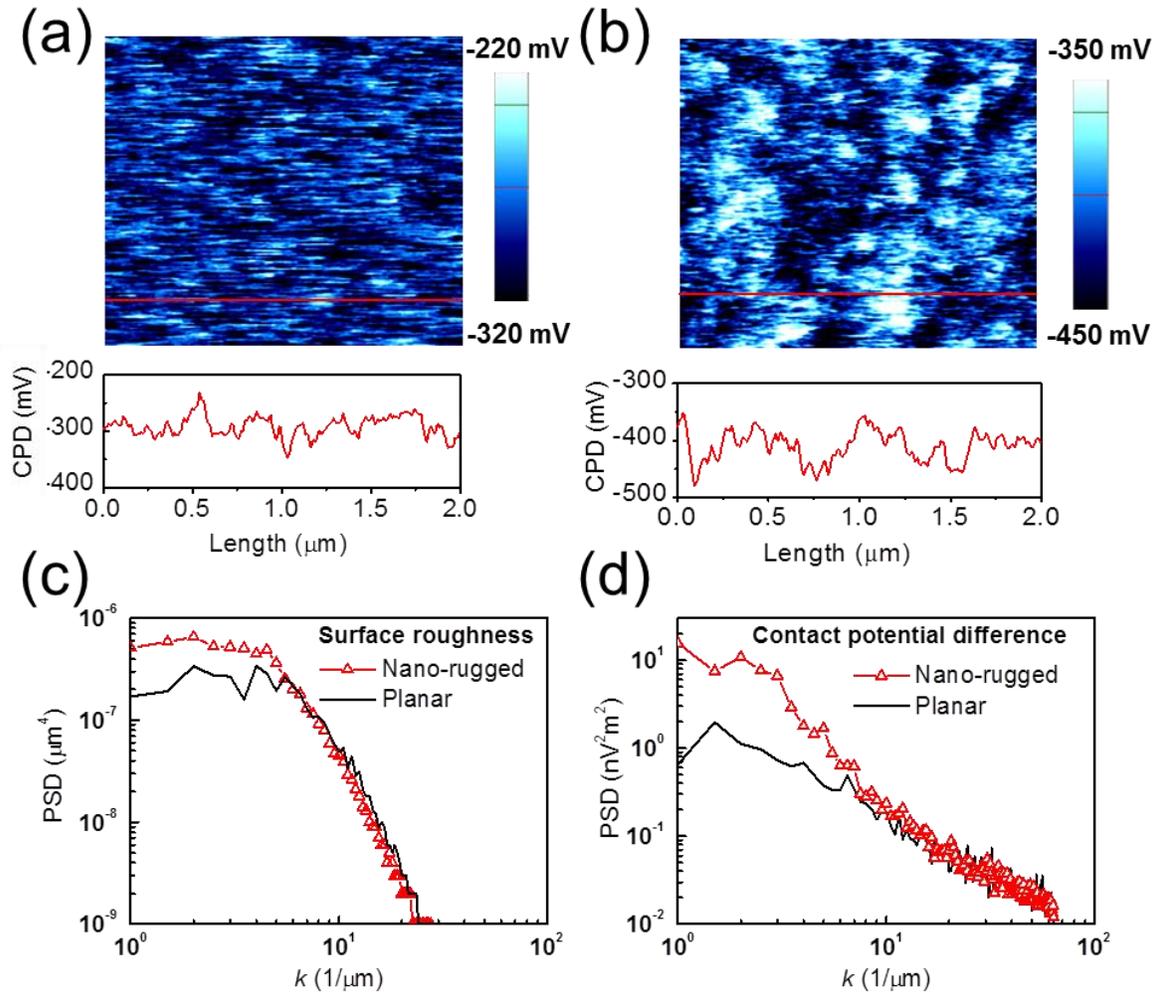

**Figure 3. Maps (scan area; 2×2 μm$^2$) and line profiles of contact potential difference (CPD) of (a) planar and (b) nano-rugged samples. The power spectral density (PSD) of (c) the surface roughness and (d) the CPD as a function of spatial frequency ($k$) of the lateral distance for the two samples.**

**Figures 3(a)** and **(b)** are maps and line profiles of the contact potential difference (CPD) of the planar and nano-rugged samples, respectively, obtained by Kelvin probe force microscopy. The CPD contrast in the planar sample seems to be originated from the grain boundaries, as revealed in the TEM images (**Figure 1**). The nano-rugged sample clearly shows the spatial modulation in the CPD values (**Figure 3b**). It should be also noted that the CPD value of the nano-rugged sample is smaller than that of the planar sample. The difference of the measured CPD data is about 0.1 eV, which is comparable to the strain-induced PL peak shift reported in literature.[23] **Figures 3(c)** and **(d)** are the plots of the power spectral density (PSD) versus the spatial frequency ($k$) of the lateral distance for the measured surface roughness and CPD, respectively. At $k < 5$ μm$^{-1}$, the PSD of the surface morphology for the samples have plateaus. The low-frequency PSD of the nano-rugged sample is larger than that of the planar sample, since the characteristic spatial frequency of the surface roughness is hundreds of nm, as shown in **Figure S5**. In the planar sample, the low-frequency PSD of the surface morphology could result from the grain size and the particle-to-particle separation. Comparison of the CPD PSD spectra of the two samples clearly reveals the differences between them. The CPD of the nano-rugged sample has much larger PSD than that of the planar counterpart at $k < 5$ μm$^{-1}$, where the large PSD of the surface roughness appears. This clearly suggests that the nano-rugged substrate affect the spatial distribution of the CPD data. Thus, the PSD analyses clearly indicate the strain effects on the MoS$_2$ layers on the nano-rugged substrates, which are consistent with the PL and Raman data (**Figure 2**).

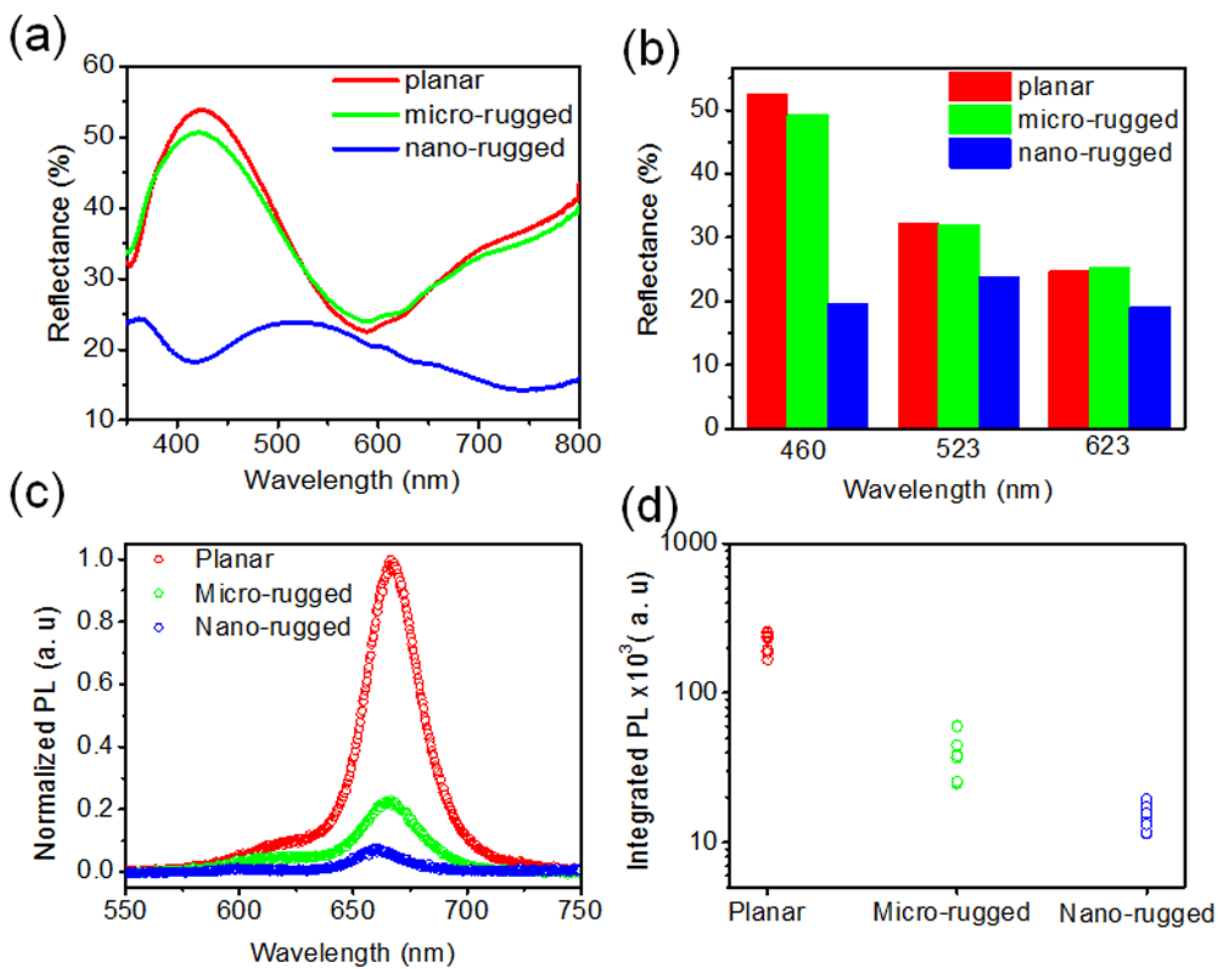

**Figure 4. (a) Reflectance spectra and (b) reflectance values at blue (~460 nm), green (~523 nm), and red (~623 nm) of the planar, micro-rugged, and nano-rugged samples. (c) Representative PL spectra and (d) spectrally integrated PL counts at 10 different spots of the planar, micro-rugged, and nano-rugged samples. In part (c), the PL spectrum from the planar sample is normalized and the others are scaled with the same factor.**

Even with enhanced light absorption compared with the planar sample, we found that the PL counts steeply drop for the rugged samples, indicative of strong PL quenching. The normalized PL spectrum from the planar sample is plotted by the red trace in **Figure 4(c)**. The properly scaled PL spectra from the other two samples are also shown to illustrate the relative PL intenstiy. Apparent blueshift of the latters again indicates that local strain is mostly compressive, which is consistent with **Figure 2**. **Figure 4(d)** shows the spectrally integrated PL counts of the three samples obtained from randomly selected 10 excitation spots. Relatively wide distributions of the PL counts observed from the rugged samples are correlated with the local strain structure that also induces wide spectral distributions as shown in **Figures 2(d)-(f)**. In contrast to tensile strain,[26, 27] on average, the intensity of *A* excitonic emission is strongly quenched by one order of magnitude for the nano-rugged sample compared to the planar sample owing to the sensitive effects of strain on the $MoS_2$ band structure, causing optical transition indirect involving the $\Sigma$-valley in the conduction band. The compressive strain induces blueshift in the *K*-valley and redshift in the $\Sigma$-valley. The net effect is blueshift of the direct transition together with the increased population in the $\Sigma$-valley, where direct transition is forbidden. The observed level of PL quenching indicates ~0.4% compressive strain.[27] Clearly, this phenomenon plays an important role in improving exciton lifetime that eventually produces about 10 times higher photocurrent.

The top and side views of our photo-response devices are schematically illustrated in **Figure S7(a)** and **(b)**, respectively. 50-nm-thick Ag electrodes were deposited on as-grown monolayer $MoS_2/SiO_2/Si$ by using thermal evaporation. A shadow mask was used to fix the distance between the two electrodes to be ~0.5 mm and the length of each electrode to be 5 mm for all samples. In order to demonstrate the device performance of our $MoS_2$ samples, we measured photocurrent generation using blue (~460 nm), green (~523 nm), and red (~623 nm) light emitting diodes (LEDs) (ModuLight, IviumStat) as light sources.

For comparison with the previous reports, the photo-responsivity, which is known as the key value for the assessment of the photodetector efficiency, was calculated using,[20]

$$R(mAW^{-1}) = \frac{I_{ph}(mA)}{P_i(Wcm^{-2}) \times A(cm^2)}$$

where $I_{ph}$ is the photocurrent generated from the device, $P_i$ is the incident radiant intensity, and $A$ is the beam area. **Figures 5(a)-(c)** illustrate the series of the photo-responsivity for the planar, micro-rugged, and nano-rugged samples as a function of input intensity under irradiation of blue (460 nm), green (523 nm), and red (623 nm) LEDs, respectively. The photo-responsivity is found to decrease with increasing the LED intensity, which is consistent with the previous results.[9, 20, 28-30] At applied bias of 0.1 V, the photo-responsivity value of the nano-rugged sample varies from ~254.5 mA/W ($P_i$ = 100 µW/cm$^2$) to ~728.0 mA/W ($P_i$ = 10 µW/cm$^2$) under irradiation of green (523 nm), which is higher than other values obtained using previous approaches in MoS$_2$-based photocurrent generation at low applied bias.[15, 19, 20] Moreover, the conformal MoS$_2$ on the nano-rugged substrate show surprisingly huge enhancement of photo-responsivity. The photo-responsivity of the nano-rugged device (~254.5 mA/W) was found to be 32 and 59 times larger compared to the micro-rugged (~7.9 mA/W) and planar (~4.3 mA/W) devices, respectively, under external applied bias of 0.1 V and input intensity of 100 µW/cm$^2$. Such a huge increase in photocurrent in the nano-rugged device could be explained by the synergistic effect of improved light absorption and strain-induced PL quenching (**Figure 4**). In other words, the rugged substrate more effectively absorbs light and generates more photo-excited carriers. The local strain in the rugged substrate induces changes in the band structure of monolayer MoS$_2$, which strongly suppresses radiative recombination of the photo-generated electron-hole pairs, enabling efficient collection of charge carriers.

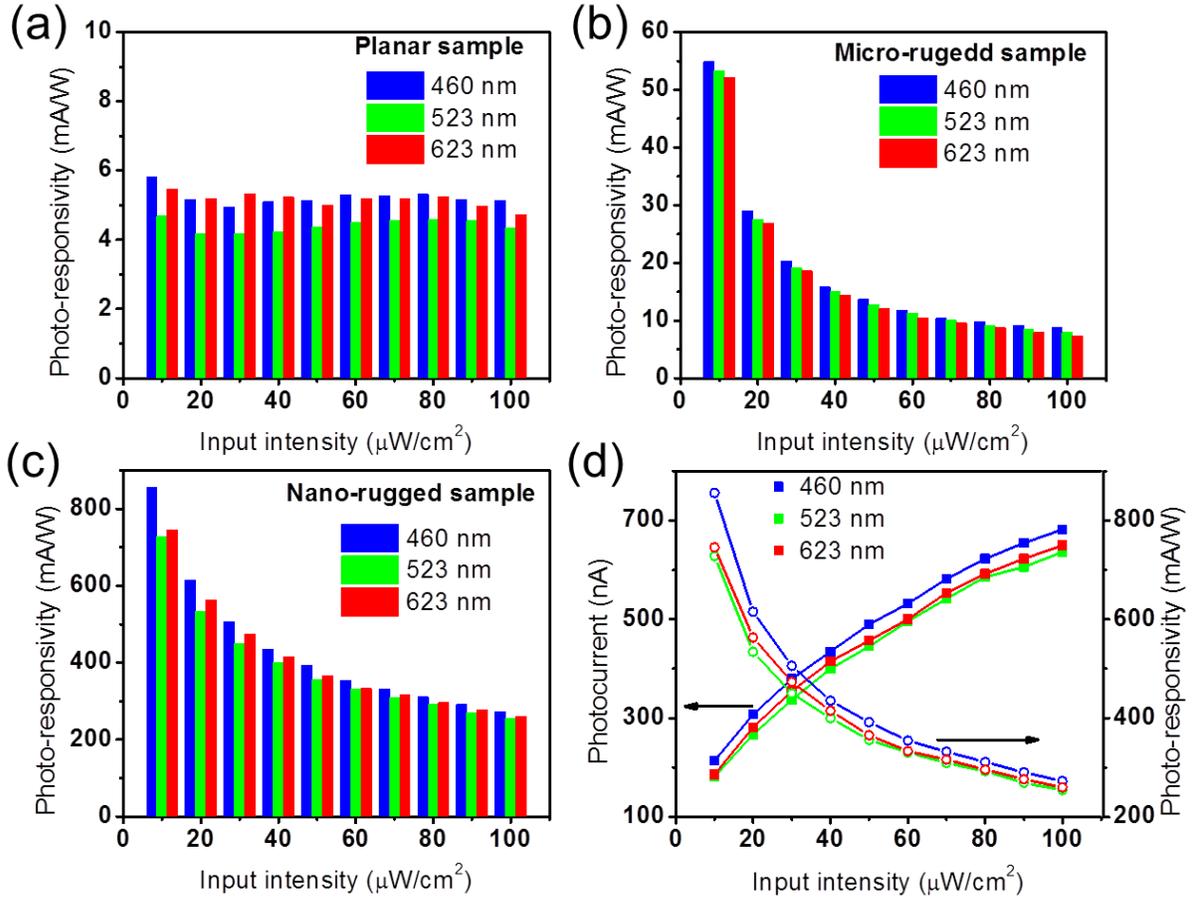

**Figure 5. The power-dependent photo-responsivity of (a) the planar, (b) micro-rugged, and (c-d) nano-rugged samples at $V_{DS}$ = 0.1 V**

We have successfully synthesized conformal monolayer MoS$_2$ on large-scale planar, micro-, and nano-rugged substrates by interconnecting neighboring nanoflakes using MOCVD under careful growth control. High-quality and uniformity of our samples were confirmed using several microscopic and spectroscopic characterization tools. More surprisingly, the fabricated photocurrent-generation device based on the nano-rugged surface exhibited 59 times higher photo-responsivity in comparison with that fabricated on the planar substrate. Enhanced light absorption and strain-induced PL quenching cumulatively contribute to highly efficient photo-response of the nano-rugged device, where PL quenching is especially unique in our approach to achieve a record-high photo-responsivity. Therefore, we believe that the large-scale conformal growth of monolayer $MoS_2$, especially on nano-rugged substrates, can be potentially employed for the batch fabrication of $MoS_2$-based atomic-thick photo-switching and photocurrent generation devices without using significant gate bias, gas adsorbates, plasmonic nanoparticles, or other accessories.

## Experimental Method

*Large-scale conformal growth of monolayer $MoS_2$*: The synthesis setup for atomic-thick $MoS_2$ is illustrated in **Figure S1**. A hot-wall furnace (CVD system, Dada) with an inner diameter of ~4 cm was used as the reaction zone where molybdenum hexacarbonyl (MHC: chemical precursor for Mo), diethyl sulfide (DES: chemical precursor for S), argon (Ar: carrier gas) and hydrogen ($H_2$: reduction gas) react to grow monolayer $MoS_2$. The MHC chamber was kept at room temperature while the DES chamber was heated to ~50 $^o$C. For cooling purpose, an Ar gas line was connected ahead of a mass flow controller of DES with a flow rate of 0.1 sccm. The MHC carrier line was gently heated at 50 $^o$C to guarantee that the Mo precursor does not remain after completing the reaction. The optimal growth parameters were set as follow: The reaction temperature is ~600 $^o$C; the reaction pressure is ~60 torr; the reaction time is 8 h; the flow rates for carrier gas (Ar), reduction gas ($H_2$) and DES are ~30 sccm, ~3 sccm, and ~1.5 sccm, respectively, whereas MHC freely flows. The planar, micro-, nano-rugged SiO$_2$/Si substrates were horizontally located at the center of the reaction chamber.

*Characterizations*: The morphologies of samples were examined using field emission electron microscopy (FE-SEM; JSM6500F, JEOL), atomic force microscopy (AFM; MFP-3D, Asylum Research), and high-resolution Cs-corrected scanning mode transmission electron microscopy (Cs-corrected STEM; ARM200F, JEM). Photoluminescence and Raman maps and spectra were taken using a micro-PL/Raman (MonoRa500i, Dongwon Optron) with an excited wavelength of ~473 nm, output power of ~5 mW, and scan step of ~0.4 μm. The contact potential difference (CPD) of samples were obtained by using Kelvin probe force microscopy (KPFM; XE-100, Park System) working in a nitrogen glove box at 20 $^{o}$C. The employed KPFM tip was a non-contact probe NSG10/Pt with PtIr conductive and reflective coating. Reflectance spectra were measured using a FT-UV-VIS-NIR microscope (Vertex 80, Bruker) with a GaP diode detector (33,000~18,000 cm$^{-1}$) and a Si diode detector (20,000~8500 cm$^{-1}$).

## Acknowledgements

This research was supported by the National Research Foundation of Korea (Nos. 2009-0093818; 2014R1A4A1071686; 2017R1E1A1A01075350; 2017R1D1A1B03035539; 2017R1A2B4005480; 2016R1D1A1A09917491). We would like to thank Dr. Bien Cuong Tran Khac for his contribution on AFM characterizations and Professor Kwanpyo Kim for his effort on measuring cross-sectional HR-TEM.

# Supporting Information

# Large-Scale Conformal Growth of Atomic-Thick MoS$_2$ for Highly Efficient Photocurrent Generation

Tri Khoa Nguyen,[1] Anh Duc Nguyen,[1] Chinh Tam Le,[1] Farman Ullah,[1] Kyo-in Koo,[2] Eunah Kim,[3] Dong-Wook Kim,[3]* Joon I. Jang,[4]* Yong Soo Kim[1]*

[1]*Department of Physics and Energy Harvest Storage Research Center (EHSRC), University of Ulsan, Ulsan 44610, South Korea*

[2]*Department of Biomedical Engineering, University of Ulsan, Ulsan 44610, South Korea*

[3]*Department of Physics, Ewha Womans University, Seoul 03760, South Korea*

[4]*Department of Physics, Sogang University, Seoul 04107, South Korea*

*Corresponding authors: yskim2@ulsan.ac.kr (Y.S. Kim), dwkim@ewha.ac.kr (D.–W. Kim), and jjcoupling@sogang.ac.kr (J.I. Jang)

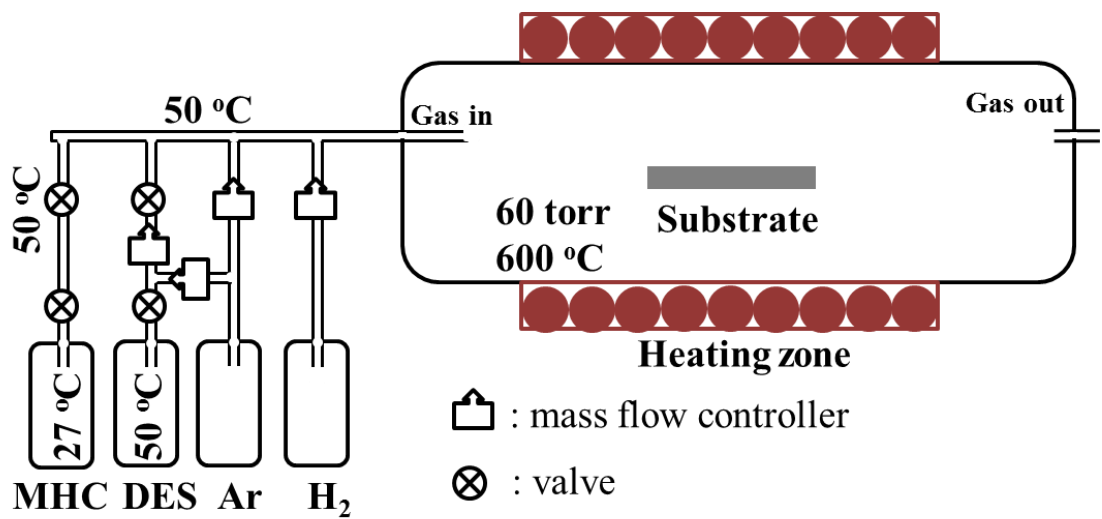

**Figure S1.** Schematic illustration of the MOCVD set up.

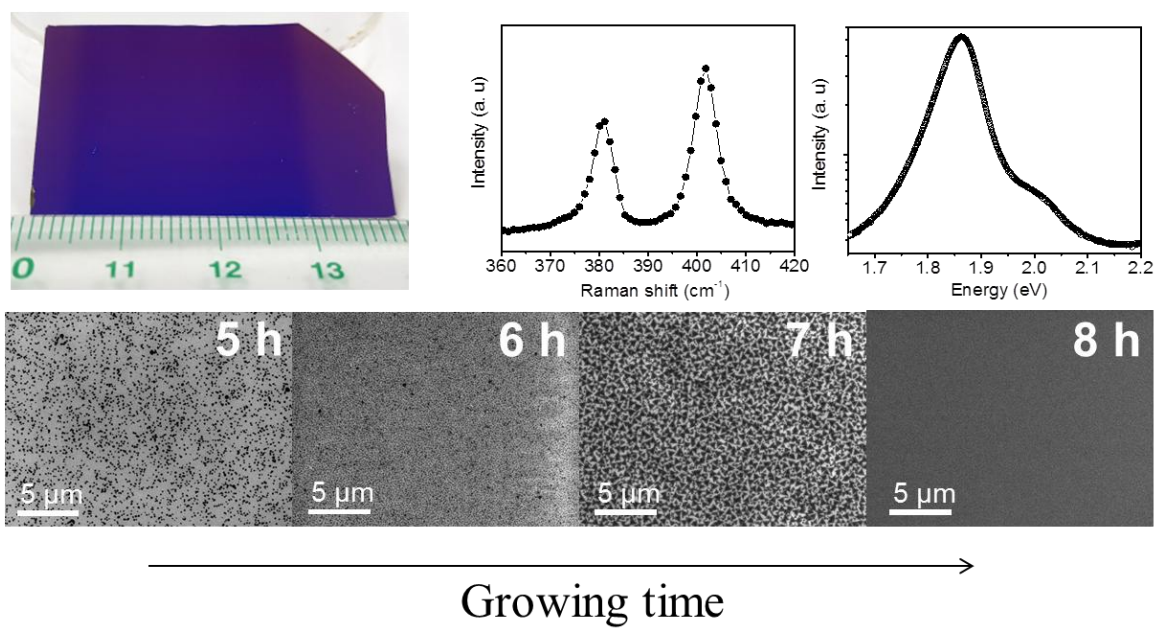

**Figure S2.** (Top) A photograph, Raman, and PL results of continuous monolayer $MoS_2$ on a planar $SiO_2/Si$ substrate. (Bottom) Corresponding morphologies of monolayer $MoS_2$ as a function of reaction time.

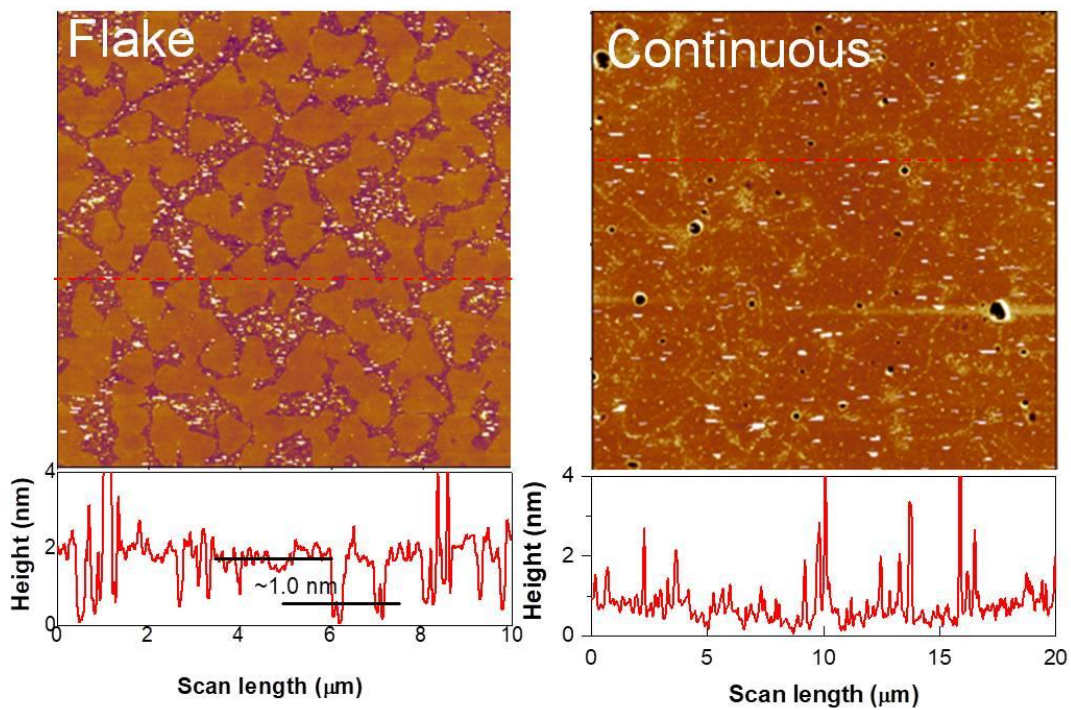

**Figure S3. AFM images (scan area; 20×20 μm$^2$) of monolayer MoS$_2$ having isolated grains (flake) and an almost fully connected region (continuous): Some minor voids are observed, which may be removed with a better growth control. Note that there is no nucleation of a second layer observed before forming continuous monolayer MoS$_2$. The grain boundaries with a thickness of ~1- 4 nm in continuous MoS$_2$ may be generated from the interconnection of neighboring flakes with different orientations. More detail is discussed in the main text.**

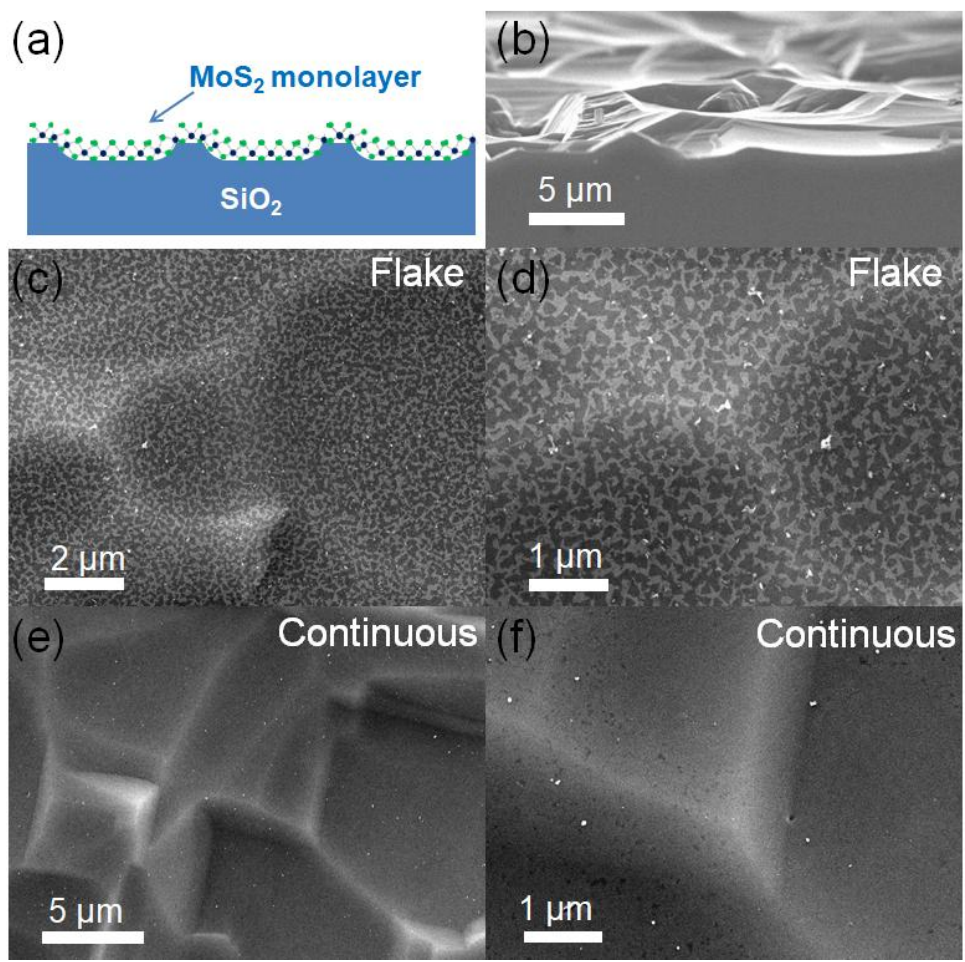

**Figure S4.** (a) Schematic illustration of monolayer MoS$_2$ grown on a micro-rugged substrate, (b) cross-section FE-SEM image of the micro-rugged substrate, and (c)-(f) FE-SEM images of monolayer MoS$_2$ grown on the micro-rugged substrate, showing the evolution from (c)-(d) nanoflakes to (e)-(f) continuous monolayer as the growth times increases.

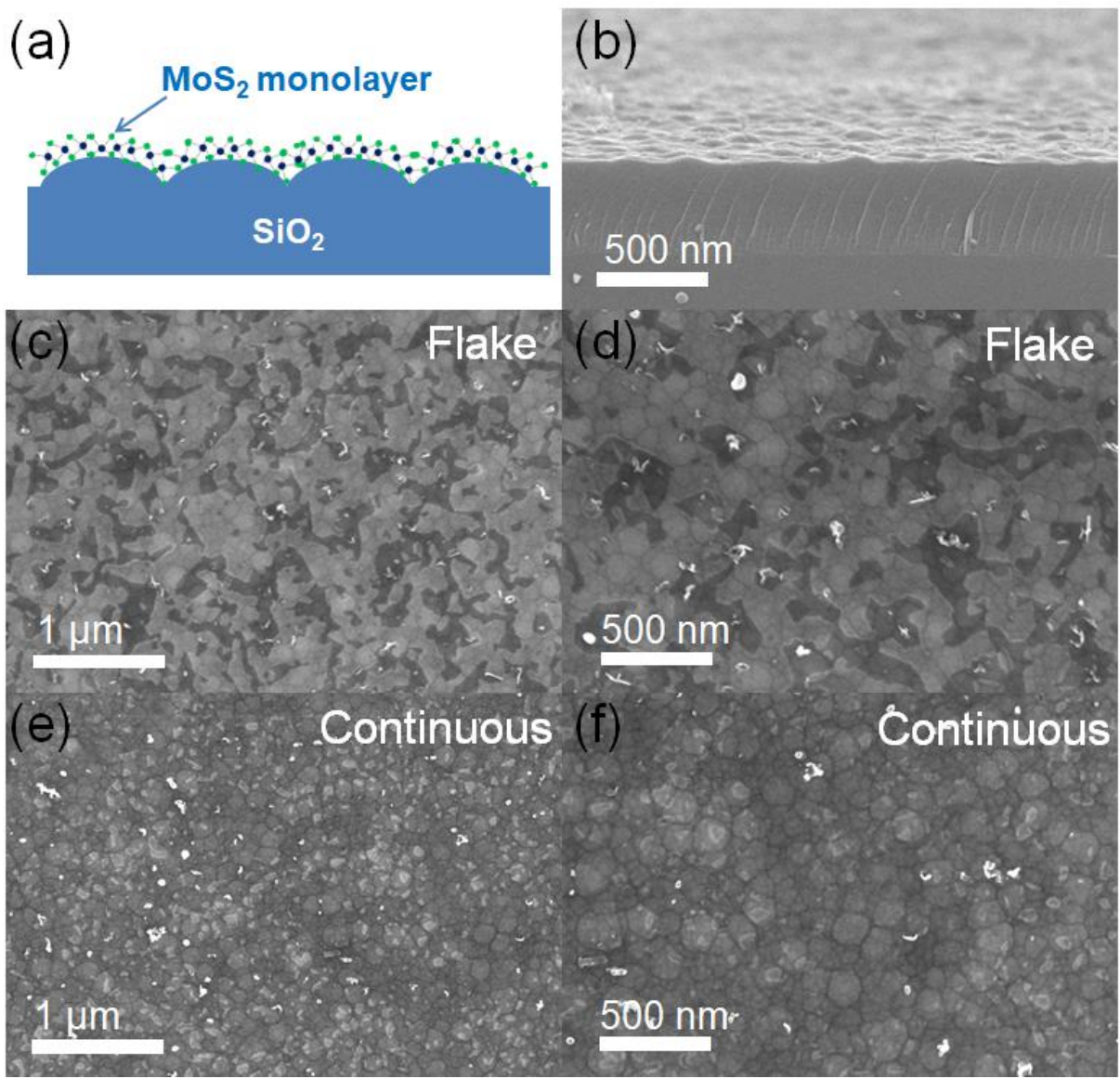

**Figure S5. (a)** Schematic illustration of monolayer MoS$_2$ grown on a nano-rugged substrate, **(b)** cross-section FE-SEM image of the nano-rugged substrate, and **(c)-(f)** FE-SEM images of monolayer MoS$_2$ grown on the nano-rugged substrate, showing the evolution from **(c)-(d)** nanoflakes to **(e)-(f)** continuous monolayer as the growth times increases.

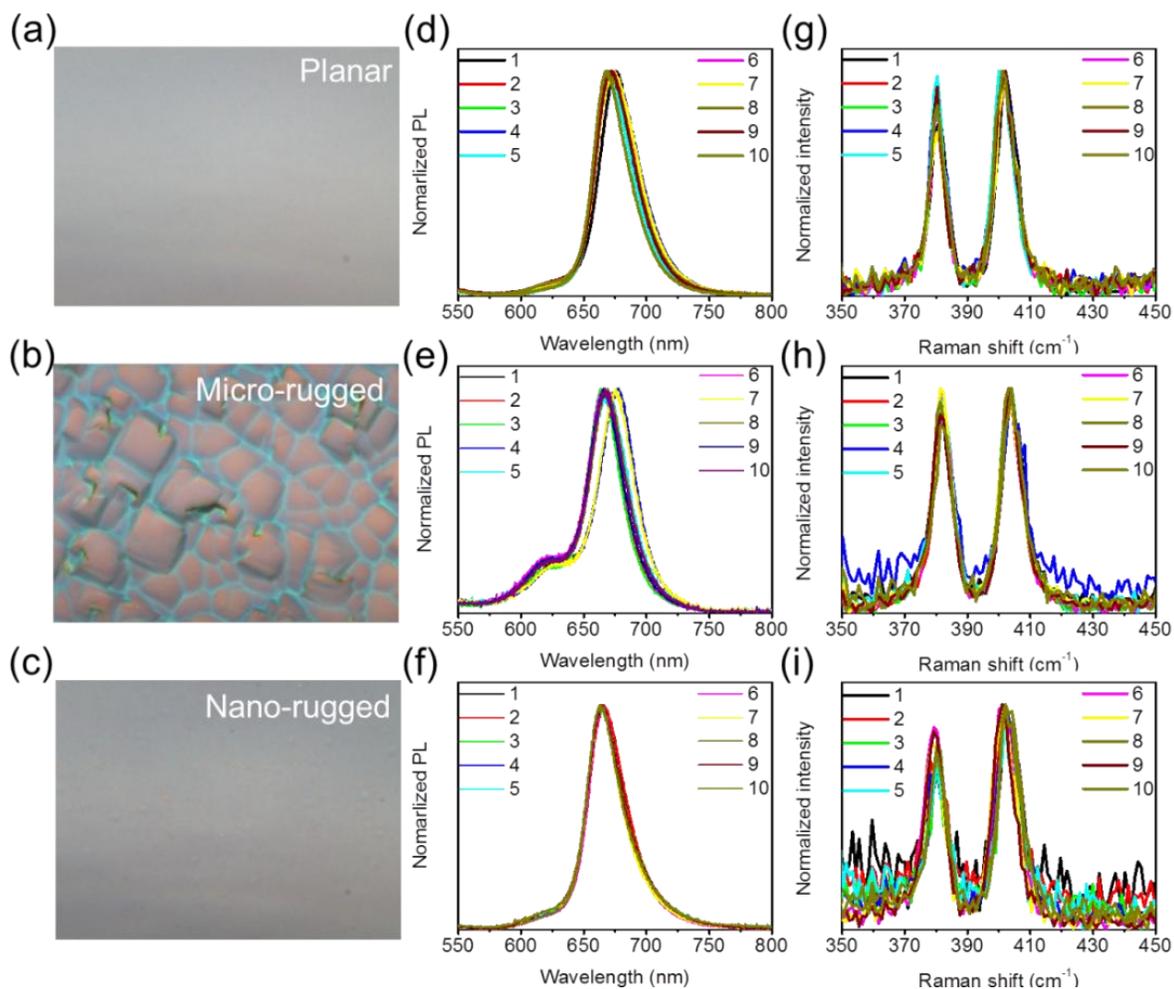

**Figure S6.** (a)-(c) Optical images of large-scale conformal atomic-thick MoS$_2$ grown on the planar, micro-rugged and nano-rugged substrates. (d)-(f) PL and (g)-(i) Raman spectra taken from 10 random positions of the planar, micro-rugged and nano-rugged samples.

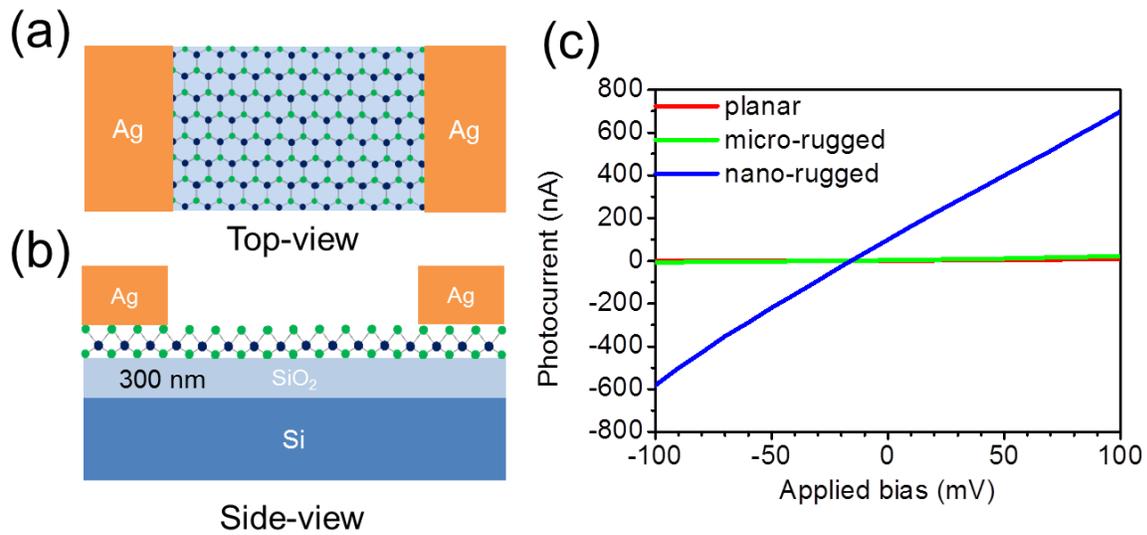

**Figure S7.** (a) Top-viewed and (b) Side-viewed schematic illustration of continuous $MoS_2$-based photo-response devices. (c) Photocurrent (excited wavelength of 460 nm and input power of 100 $\mu W/cm^2$) of samples under applied bias from -100 to 100 mV.